\documentclass[longauth]{aa} 
\usepackage{graphicx,natbib}
\usepackage{txfonts}
\begin{document}
   \title{Another look at the \object{BL Lacertae} flux and spectral variability} 

   \subtitle{Observations by GASP-WEBT, XMM-Newton, and Swift in 2008--2009\thanks{The radio-to-optical 
   data presented in this paper are stored in the GASP-WEBT archive; 
   for questions regarding their availability,
   please contact the WEBT President Massimo Villata ({\tt villata@oato.inaf.it}).}}

   \author{C.~M.~Raiteri              \inst{ 1}
   \and   M.~Villata                  \inst{ 1}
   \and   L.~Bruschini                \inst{ 2}
   \and   A.~Capetti                  \inst{ 1}
   \and   O.~M.~Kurtanidze            \inst{ 3}
   \and   V.~M.~Larionov              \inst{ 4,5,6}
   \and   P.~Romano                   \inst{ 7}
   \and   S.~Vercellone               \inst{ 7}
   \and   I.~Agudo                    \inst{ 8,9}
   \and   H.~D.~Aller                 \inst{10}
   \and   M.~F.~Aller                 \inst{10}
   \and   A.~A.~Arkharov              \inst{ 5}
   \and   U.~Bach                     \inst{11}
   \and   A.~Berduygin                \inst{12}
   \and   D.~A.~Blinov                \inst{ 4}
   \and   M.~B\"ottcher               \inst{13}
   \and   C.~S.~Buemi                 \inst{14}
   \and   P.~Calcidese                \inst{15}
   \and   D.~Carosati                 \inst{16}
   \and   R.~Casas                    \inst{17,18}
   \and   W.-P.~Chen                  \inst{19}
   \and   J.~Coloma                   \inst{20}
   \and   C.~Diltz                    \inst{13}
   \and   A.~Di Paola                 \inst{21}
   \and   M.~Dolci                    \inst{22}
   \and   N.~V.~Efimova               \inst{ 4,5}
   \and   E.~Forn\'e                  \inst{18}
   \and   J.~L.~G\'omez               \inst{ 9}
   \and   M.~A.~Gurwell               \inst{23}
   \and   A.~Hakola                   \inst{12}
   \and   T.~Hovatta                  \inst{24,25}
   \and   H.~Y.~Hsiao                 \inst{19,26}
   \and   B.~Jordan                   \inst{27}
   \and   S.~G.~Jorstad               \inst{ 8}
   \and   E.~Koptelova                \inst{19}
   \and   S.~O.~Kurtanidze            \inst{ 3}
   \and   A.~L\"ahteenm\"aki          \inst{24}
   \and   E.~G.~Larionova             \inst{ 4}
   \and   P.~Leto                     \inst{14}
   \and   E.~Lindfors                 \inst{12}
   \and   R.~Ligustri                 \inst{28}
   \and   A.~P.~Marscher              \inst{ 8}
   \and   D.~A.~Morozova              \inst{ 4}
   \and   M.~G.~Nikolashvili          \inst{ 3}
   \and   K.~Nilsson                  \inst{29}
   \and   J.~A.~Ros                   \inst{18}
   \and   P.~Roustazadeh              \inst{13}
   \and   A.~C.~Sadun                 \inst{30}
   \and   A.~Sillanp\"a\"a            \inst{12}
   \and   J.~Sainio                   \inst{12}
   \and   L.~O.~Takalo                \inst{12}
   \and   M.~Tornikoski               \inst{24}
   \and   C.~Trigilio                 \inst{14}
   \and   I.~S.~Troitsky              \inst{ 4}
   \and   G.~Umana                    \inst{14}
 }

   \offprints{C.~M.~Raiteri}

   \institute{
          INAF, Osservatorio Astronomico di Torino, Italy                                                     
   \and   Dipartimento di Fisica Generale, Universit\`a di Torino, Italy                                      
   \and   Abastumani Observatory, Mt. Kanobili, Georgia                                                       
   \and   Astron.\ Inst., St.-Petersburg State Univ., Russia                                                  
   \and   Pulkovo Observatory, St.\ Petersburg, Russia                                                        
   \and   Isaac Newton Institute of Chile, St.-Petersburg Branch                                              
   \and   INAF-IASF Palermo, Italy                                                                            
   \and   Institute for Astrophysical Research, Boston University, MA, USA                                    
   \and   Instituto de Astrof\'{i}sica de Andaluc\'{i}a, CSIC, Granada, Spain                                 
   \and   Department of Astronomy, University of Michigan, MI, USA                                            
   \and   Max-Planck-Institut f\"ur Radioastronomie, Bonn, Germany                                            
   \and   Tuorla Observatory, Dept.\ of Physics and Astronomy, Univ.\ of Turku, Finland                       
   \and   Department of Physics and Astronomy, Ohio Univ., OH, USA                                            
   \and   INAF, Osservatorio Astrofisico di Catania, Italy                                                    
   \and   Osservatorio Astronomico della Regione Autonoma Valle d'Aosta, Italy                                
   \and   Armenzano Astronomical Observatory, Italy                                                           
   \and   Inst.\ de Ci\`encies de l'Espai (CSIC-IEEC), Spain                                                  
   \and   Agrupaci\'o Astron\`omica de Sabadell, Spain                                                        
   \and   Graduate Institute of Astronomy, National Central University, Taiwan                                
   \and   Observatori El Vendrell, Spain                                                                      
   \and   INAF, Osservatorio Astronomico di Roma, Italy                                                       
   \and   INAF, Osservatorio Astronomico di Collurania Teramo, Italy                                          
   \and   Harvard-Smithsonian Center for Astrophysics, Cambridge, MA, USA                                     
   \and   Aalto University Mets\"ahovi Radio Observatory, Finland                                             
   \and   Department of Physics, Purdue University, USA                                                       
   \and   Lulin Observatory, National Central University, Taiwan                                              
   \and   School of Cosmic Physics, Dublin Institute For Advanced Studies, Ireland                            
   \and   Circolo Astrofili Talmassons, Italy                                                                 
   \and   Finnish Centre for Astronomy with ESO (FINCA), University of Turku, Finland                         
   \and   Dept.\ of Phys., Univ.\ of Colorado Denver, CO, USA                                                 
 }

   \date{}
 
  \abstract
   {}
   {In a previous study we suggested that the broad-band emission and variability properties of BL Lacertae can be accounted for by a double synchrotron emission component with related inverse-Compton emission from the jet, plus thermal radiation from the accretion disc. Here we investigate the matter with further data extending over a wider energy range.} 
   {The GLAST-AGILE Support Program (GASP) of the Whole Earth Blazar Telescope (WEBT) monitored BL Lacertae in 2008--2009 at radio, near-IR, and optical frequencies to follow its flux behaviour. During this period, high-energy observations were performed by XMM-Newton, Swift, and Fermi. We analyse these data with particular attention to the calibration of Swift UV data, and apply a helical jet model to interpret the source broad-band variability.}
   {The GASP-WEBT observations show an optical flare in 2008 February--March, and oscillations of several tenths of mag on a few-day time scale afterwards. The radio flux is only mildly variable. The UV data from both XMM-Newton and Swift seem to confirm a UV excess that is likely caused by thermal emission from the accretion disc. The X-ray data from XMM-Newton indicate a strongly concave spectrum, as well as moderate ($\sim 4$--7\%) flux variability on an hour time scale. The Swift X-ray data reveal fast (interday) flux changes, not correlated with those observed at lower energies. We compare the spectral energy distribution (SED) corresponding to the 2008 low-brightness state, which was characterised by a synchrotron dominance, to the 1997 outburst state, where the inverse-Compton emission was prevailing. A fit with an inhomogeneous helical jet model suggests that two synchrotron components are at work with their self inverse-Compton emission. Most likely, they represent the radiation from two distinct emitting regions in the jet. We show that the difference between the source SEDs in 2008 and 1997 can be explained in terms of pure geometrical variations. The outburst state occurred when the jet-emitting regions were better aligned with the line of sight, producing an increase of the Doppler beaming factor.}
  {Our analysis demonstrates that the jet geometry can play an extremely important role in the BL Lacertae flux and spectral variability.
Indeed, the emitting jet is probably a bent and dynamic structure, and hence changes in the emitting regions viewing angles are likely to happen, with strong consequences on the source multiwavelength behaviour.}



   \maketitle
%

   \begin{figure*}
   \sidecaption
   \includegraphics[width=14cm]{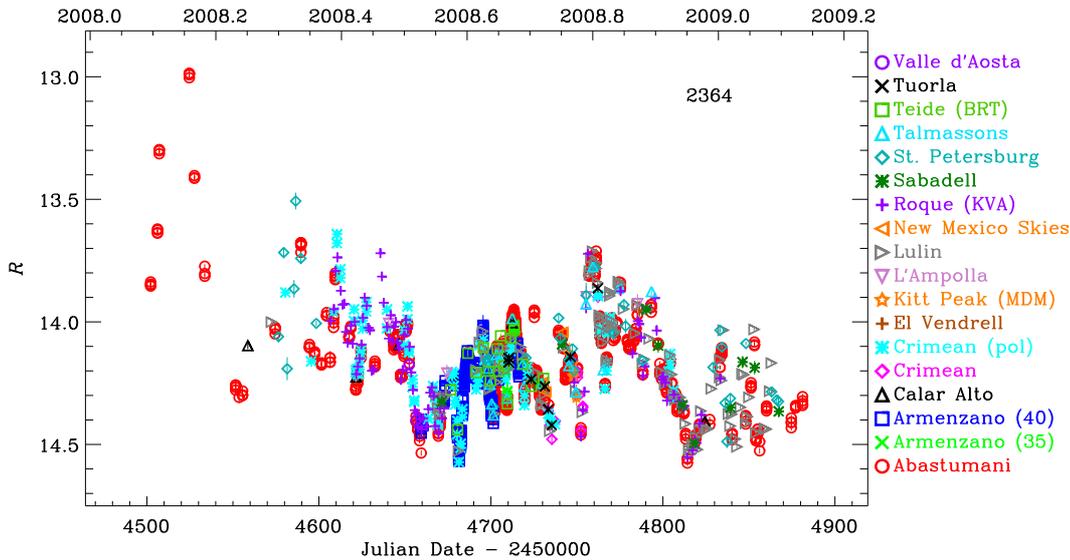}
      \caption{$R$-band light curve by the GASP collaboration 
from February 2008 to February 2009. Data are not corrected for the host galaxy contamination or Galactic extinction. The participating observatories are marked 
with different symbols and colours. The total number of data points is indicated in the upper right.} 
         \label{erre}
   \end{figure*}

\section{Introduction}

Blazars are active galactic nuclei whose extreme properties are thought to be owing to their relativistic jets pointing toward us.
BL Lacertae, the prototype of the ``BL Lac objects" blazar class,
has been the target of many campaigns by the Whole Earth Blazar Telescope (WEBT) 
collaboration\footnote{{\tt http://www.oato.inaf.it/blazars/webt/}} since 1999. 
The tens of thousands of optical-to-radio data collected by the WEBT allowed us to study its multiwavelength flux variability, colour behaviour, and the correlations among flux variations in different bands, and revealed a possible periodicity of the radio outbursts. The results have been published by \citet{vil02,rav02,boe03,vil04a,vil04b,bac06,pap07,vil09a,lar10}.

In a recent paper, \citet{rai09} analysed the multiwavelength data from the 2007--2008 WEBT campaign, including three pointings by XMM-Newton. 
The XMM-Newton data revealed a UV excess, which was interpreted to be due to thermal emission from the accretion disc, as well as a spectral curvature in the X-ray band.
The authors constructed spectral energy distributions (SEDs) of BL Lacertae corresponding to various epochs where the source was in different brightness states, using both their own data and data from the literature.
They applied the inhomogeneous, rotating helical jet model by \citet[][see also \citealt{rai99}, \citealt{rai03}, \citealt{ost04}] {vil99} to fit the SEDs, and 
suggested that the broad-band spectral properties of BL Lacertae may result from the combination of two synchrotron emission components with their self inverse-Compton emission, plus a thermal component from the disc.
Subsequently, \citet{cap10} analysed optical spectra acquired in the same period with the 3.56 m Telescopio Nazionale Galileo (TNG). They found a broad H$\alpha$ 
emission line, with luminosity of $\sim 4 \times 10^{41} \, \rm erg \, s^{-1}$ and FWHM of $\sim 4600 \rm \, km \, s^{-1}$, even brighter than that found in 1995--1997 by \citet{ver95} and \citet{cor96,cor00}. This favours the hypothesis that the UV excess is caused by thermal emission from the accretion disc, the most likely source of ionising photons for the broad line region. 
The multiwavelength data available for the \citet{rai09} analysis lacked simultaneous information in the $\gamma$-ray band, so that the inverse-Compton spectral region was poorly constrained. But in 2008 the Fermi satellite was able to detect BL Lacertae \citep{abd10_sed}, even if in a low state compared to the past detections by the Compton Gamma Ray Observatory\footnote{\tt http://heasarc.gsfc.nasa.gov/docs/cgro/} (CGRO, \citealt{har99}, \citealt{blo97}). In the same period, observations in the UV and X-ray bands were performed by Swift, while in the optical, near-IR, mm and cm radio bands the source was monitored by the GLAST-AGILE Support Program (GASP) of the WEBT.
This offered the unique opportunity to study the source emission over a very extended spectral range.
The results of this further investigation effort on BL Lacertae are presented in this paper.

\section{GASP observations}
\label{sec_gasp}

   \begin{figure*}
   \sidecaption
   \includegraphics[width=12cm]{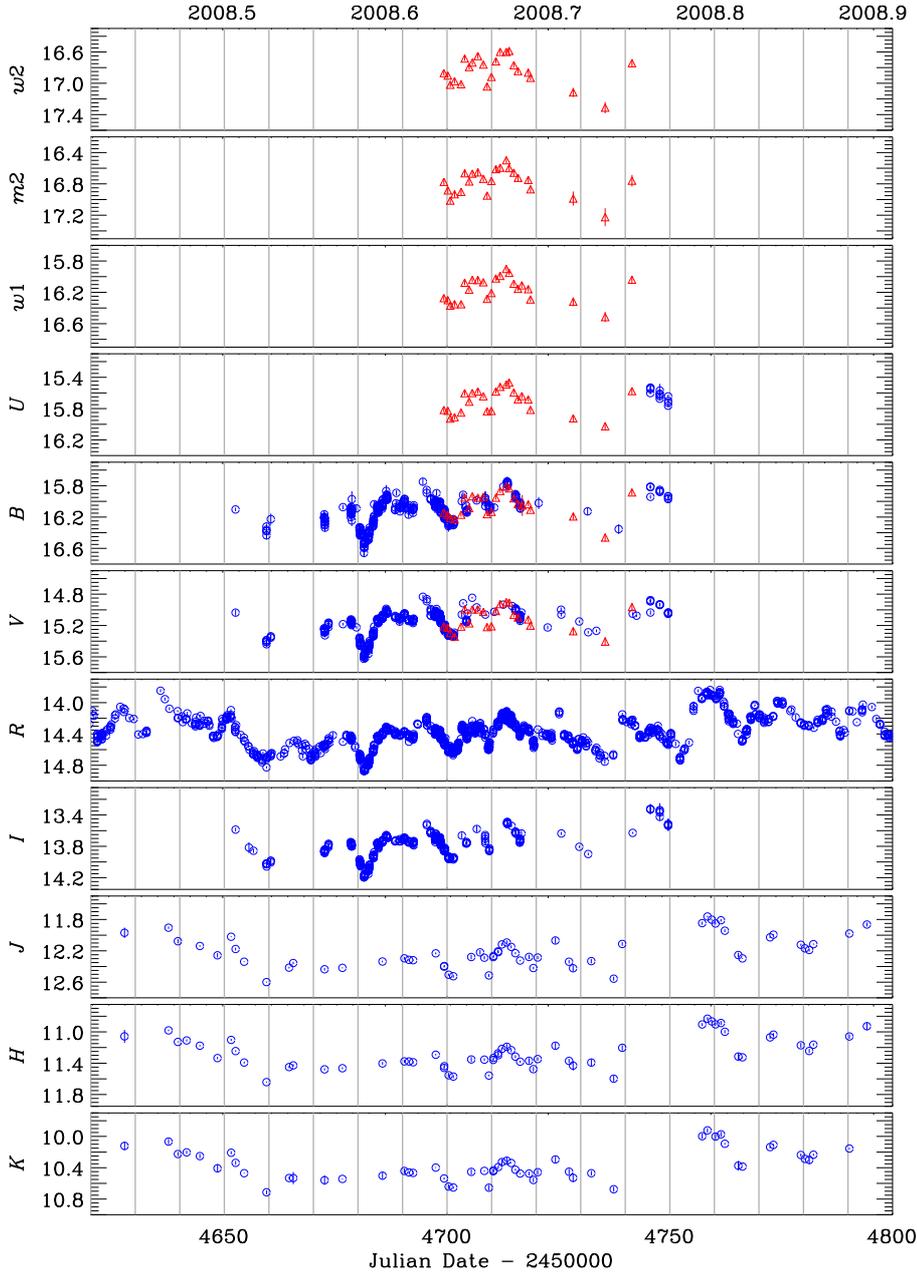}
      \caption{June--November 2008 light curves of BL Lacertae from UV to near-IR after correction for the host galaxy contribution, but not for the Galactic extinction. Data taken by the GASP-WEBT collaboration are plotted as blue circles, UVOT data as red triangles. The UVOT $u$, $b$, and $v$ light curves have been shifted to match the ground-based $U$, $B$, and $V$ ones (see text for details). }
         \label{uvir}
   \end{figure*}

The GASP was born in 2007 as a WEBT project, with the aim of monitoring a list of 28 $\gamma$-ray loud blazars in the optical, near-IR, mm, and cm radio bands during the $\gamma$-ray observations of the AGILE\footnote{\tt http://agile.iasf-roma.inaf.it/} and Fermi\footnote{\tt http://fermi.gsfc.nasa.gov/} (formerly GLAST) satellites \citep[see e.g.][]{vil08,vil09b}.
Data are collected periodically by the WEBT President, who checks the consistency of the various datasets. The GASP light curves are then available for multiwavelength studies, mostly in the framework of the GASP collaboration with the AGILE and Fermi research teams.
The GASP data presented in this paper were taken at the observatories listed in Table \ref{obs}.

\begin{table}
\caption{List of optical, near-IR, and mm--cm radio observatories contributing data to this work.}
\label{obs}
\centering
\begin{tabular}{l r c  }
\hline\hline
\multicolumn{3}{c}{\it Optical and near-infrared}\\
Observatory    & Tel.\ size    & Bands\\
               & [cm]          &      \\
\hline
Abastumani, Georgia      &  70         & $R$                \\
Armenzano, Italy         &  35         & $BRI$              \\
Armenzano, Italy         &  40         & $BVRI$             \\
Calar Alto, Spain$^a$    & 220         & $R$                \\
Campo Imperatore, Italy  & 110         & $JHK$          \\
Crimean, Ukraine         &  70         & $BVRI$             \\
El Vendrell, Spain       &  20         & $R$                \\
Kitt Peak (MDM), USA     & 130         & $UBVRI$            \\
L'Ampolla, Spain         &  36         & $R$                \\
Lulin, Taiwan            &  40         & $R$                \\
New Mexico Skies, USA    &  30         & $VRI$              \\
Roque (KVA), Spain       &  35         & $R$                \\
Sabadell, Spain          &  50         & $R$                \\
St.\ Petersburg, Russia  &  40         & $BVRI$             \\
Talmassons, Italy        &  35         & $BVR$              \\
Teide (BRT), Spain       &  35         & $BVR$              \\
Tuorla, Finland          & 103         & $R$                \\
Valle d'Aosta, Italy     &  81         & $BVRI$             \\
\hline
\multicolumn{3}{c}{\it Radio}\\
Observatory    & Tel.\ size    & Frequencies\\
               & [m]           &   [GHz]   \\
\hline
Mauna Kea (SMA), USA     &$8 \times 6^b$ & 230, 345       \\
Medicina, Italy          & 32            & 5, 8, 22         \\
Mets\"ahovi, Finland     & 14            & 37              \\
Noto, Italy              & 32            & 43             \\
UMRAO, USA               & 26            & 4.8, 8.0, 14.5   \\
\hline
\multicolumn{3}{l}{$^a$ Calar Alto data were acquired as part of the MAPCAT (Monitoring}\\
\multicolumn{3}{l}{AGN with Polarimetry at the Calar Alto Telescopes) project.}\\
\multicolumn{3}{l}{$^b$ Radio interferometer including 8 dishes of 6 m size.}
\end{tabular}
\end{table}

The optical data were calibrated with respect to a common choice of reference stars in the same field of the source (\citealt{ber69} in $U$ and $B$ bands; \citealt{fio96} in $V$, $R$, and $I$).
The source photometry was evaluated from a circular region with an 8 arcsec aperture radius, while the background was taken in a surrounding annulus with 10 and 16 arcsec radii. In this way the measure is essentially seeing-independent and all datasets are affected by the same contamination from the light of the host galaxy. \citet{rai09} estimated that with the above prescriptions the contamination amounts to about 60\% of the host total flux density, which is 0.36, 1.30, 2.89, 4.23, 5.90, 11.83, 13.97, and 10.62 mJy in the $U$, $B$, $V$, $R$, $I$, $J$, $H$, and $K$ bands, respectively.
When converting magnitudes into flux densities, we corrected for the Galactic extinction according to the \citet{car89} laws, using $R_V=3.1$, the standard value for the diffuse interstellar medium, and $A_B=1.42$ \citep[from][]{sch98}. We adopted the absolute fluxes by \citet{bes98}.

Figure \ref{erre} shows the best-sampled total
$R$-band light curve from February 2008 to February 2009 built with GASP data, which are  
not corrected for the host galaxy contribution here.
A noticeable flare was observed at the beginning of the period, in 2008 February--March; 
afterwards both the average brightness level and the variability amplitude decreased.
However, the source remained active, its brightness oscillating by several tenths of magnitude on a few-day time scale. This is not an unusual behaviour for BL Lacertae \citep[see e.g.][who reported on a 0.9 mag brightening in 24 hours]{rai09}. 

Optical data at other wavelengths as well as near-IR data are shown in Fig.\ \ref{uvir} for the period June--November 2008 (the UV data displayed in the figure are presented in Sect.\ \ref{sec_swift}).
The host galaxy contribution has been subtracted to distinguish the behaviour of the active nucleus. 
This reveals that the brightness evolution follows the same trend in the various bands, but magnitude variations are more pronounced at higher frequencies, which is a common feature of BL Lac objects.
For example, the brightness increase following the almost symmetric dip around $\rm JD \sim 2454681$ involved a variation of 0.79, 0.64, 0.57, and 0.54 mag in $\sim 5$ days in the $B$, $V$, $R$, and $I$ bands, respectively. We notice that the source redshift is $z=0.0686$ \citep{ver95} and hence the broad H$\alpha$ emission line enters the tails of the $R$ and $I$ passbands. However, referring to \citet{cap10}, one can estimate that its flux contribution is only a few thousandths of that of the continuum. Hence, the presence of the line cannot affect the variability in these bands, at least when the source brightness is at these levels.
Another interesting example of fast variability is the rise of $\sim 0.9$ mag in 4 days, from $\rm JD = 2454752.3$ to 2454756.3, in the $R$ band, which unfortunately was not observed in other bands.

\begin{figure}
   \resizebox{\hsize}{!}{\includegraphics{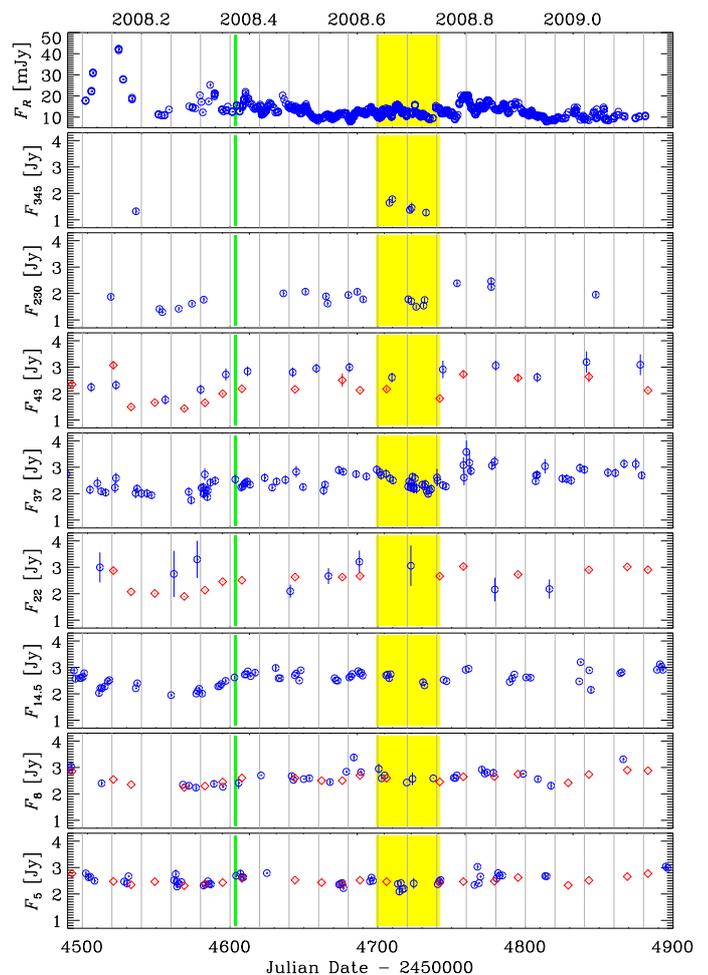}}
      \caption{Optical flux densities ($R$ band, top panel), after correction for Galactic extinction 
       and host galaxy contamination, compared to radio flux densities at different frequencies. 
       Blue circles represent GASP data; red diamonds indicate data from the VLA/VLBA PCD.
       The vertical green line corresponds to the XMM-Newton pointing of May 16--17; 
       the yellow strip highlights the period of Swift observations, from 2008 August 20 to October 2.}
         \label{radio}
   \end{figure}

Radio flux densities at cm--mm wavelengths are displayed in Fig.\ \ref{radio} together with de-reddened and host-galaxy subtracted $R$-band optical flux densities.
The former are complemented by data from the VLA/VLBA Polarization Calibration Database\footnote{\tt http://www.vla.nrao.edu/astro/calib/polar/} (PCD). 
The flux variation amplitude appears to decrease from the highest to the lowest frequencies, as usual.
One interesting feature is the fast radio flare that is visible in the 37 GHz light curve at $\rm JD \sim 2454760$, because it occurred simultaneously with an optical flare. The correlation between optical and radio flux variations in BL Lacertae has been the subject of several studies, and most of them found a correlation with a long time delay (a few months) of the radio after the optical flux changes \citep[see e.g.][]{huf92,tor94b,cle95,vil04b,bac06,vil09a}. However, simultaneous variations have already been found \citep{tor94a}. The 37 GHz data we are dealing with are affected by large uncertainties due to unfavourable weather conditions; but an increase of the radio flux is visible also at 43, 22, and 14.5 GHz, giving strength to the possibility that these events are correlated.
We also notice that there is neither a contemporaneous nor a delayed clear radio counterpart to the optical flare observed at the beginning of the period ($\rm JD \sim 2454520$).
Yet, one would have expected to see it in the high-frequency radio light curves. 
A better sampling in the mm wavelength range perhaps would have helped to understand whether the variability mechanism responsible for this flare affected the optical emission region only or if it extended also to the radio domain.

\section{Swift observations}
\label{sec_swift}
The Swift satellite observed BL Lacertae in 2008 August, September, and October, for a total of 24 pointings. In particular, from August 20 to September 9 a daily sampling was obtained.

\subsection{UVOT data}
\label{sec_uvot}
The Ultraviolet/Optical Telescope (UVOT; \citealt{rom05}) onboard the Swift spacecraft acquires data in the optical $v$, $b$, and $u$ bands, as well as in the UV filters $uvw1$, $uvm2$, and $uvw2$ \citep{poo08}.

We reduced the BL Lacertae data with the HEAsoft package version 6.7, with CALDB updated at the end of November 2009. Source counts were extracted from a circular region with a radius of 5 arcsec, while background counts were estimated in a neighbouring source-free region.
When multiple exposures in the same filter were present during an observing epoch, we first processed each frame separately with the task {\tt uvotmaghist} and then binned the results. These values were compared with those obtained by first summing the frames acquired in the same band with {\tt uvotimsum}, and then performing the aperture photometry with the task {\tt uvotsource}. We verified that the two methods are equivalent.

The final UVOT light curves are shown in Fig.\ \ref{uvir}. We subtracted the host galaxy contribution, 
taking into account that with the 5 arcsec aperture radius we used for the photometry, about 50\% of the host flux was included. We adopted the host galaxy flux densities given by \citet[][see also Sect.\ \ref{sec_gasp}]{rai09}; these authors also discussed that the host contribution can be considered negligible in the UV.

The comparison between the UVOT $u$, $b$, and $v$ data and the $U$, $B$, and $V$ data taken by the GASP observers reveals that an offset is present between the space light curves and the ground-based ones. We estimated mean offsets $U-u=0.2$, $B-b=0.1$, and $V-v=-0.05$. The UVOT light curves shown in Fig.\ \ref{uvir} have been shifted accordingly. 
Taking into account that the average UVOT colour indices of BL Lacertae are $u-b \sim -0.4$ and $b-v \sim 0.8$, the above offsets disagree with those derived by \citet{poo08} for the objects on which they based their photometric calibration of UVOT, i.e.\ Pickles stars and GRB models. Indeed, these objects have a different spectral shape, so that the \citet{poo08} calibrations may not hold for BL Lacertae.

The UVOT data confirm the variability trend traced by the ground-based ones, extending it to UV frequencies. This indicates that the variability mechanism affecting the near-IR--optical emission, which is dominated by beamed synchrotron radiation, can also produce flux changes in the UV, where a contribution from the synchrotron emission is thus expected, besides a possible contribution from thermal disc radiation.

   \begin{figure}
   \resizebox{\hsize}{!}{\includegraphics{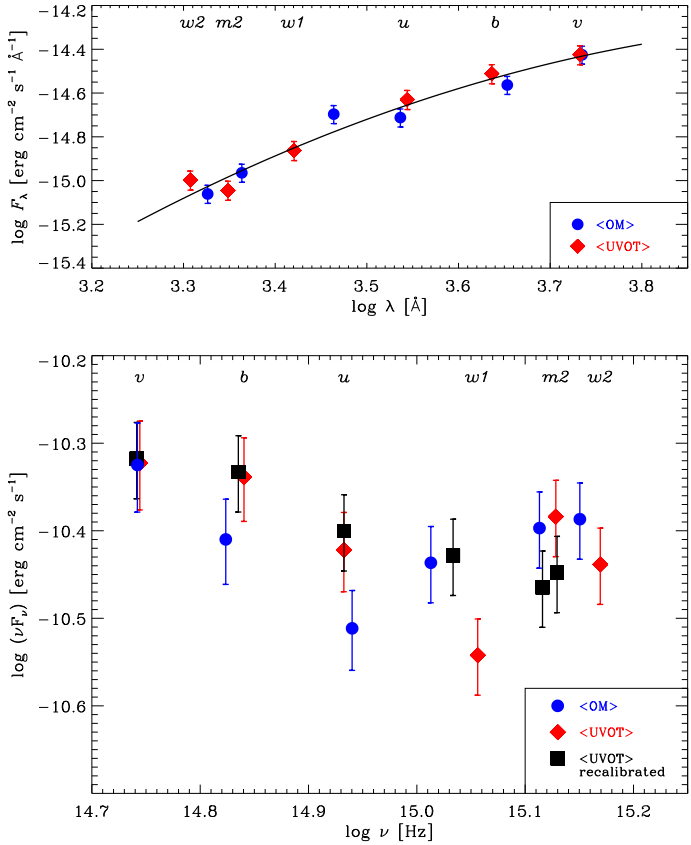}}
      \caption{Top: Observed spectrum of BL Lacertae in the optical--UV band. 
Red diamonds refer to the mean UVOT spectrum resulting from 16 epochs analysed in this paper.
Blue circles represent the average spectrum obtained from the three observations of the OM instrument onboard XMM-Newton in 2007--2008 \citep{rai09}, normalised to the mean UVOT spectrum in the $v$ band. 
The solid line is the log-parabolic fit used in the calibration procedure.
Bottom: Optical--UV SEDs of BL Lacertae. Blue circles and red diamonds are derived from the OM and UVOT average spectra shown in the top panel by using standard prescriptions to obtain dereddened flux densities. Black squares represent the mean UVOT SED after recalibration as explained in the text.
In both panels the filter labels are centred on the standard UVOT $\lambda_{\rm eff}$ \citep{poo08}.}
         \label{calib}
   \end{figure}

The derivation of the source intrinsic flux densities for further analysis (see Sect.\ \ref{sec_model}) requires some attention. In their paper on the photometric calibration of UVOT, \citet{poo08} give effective wavelengths of 5402, 4329, 3501, 2634, 2231, and 2030 \AA\ for the $v$, $b$, $u$, $uvw1$, $uvm2$, and $uvw2$ filters, respectively, but warn that the $\lambda_{\rm eff}$ of the UV filters will be longer for very red spectra. 
Moreover, they provide count-rate-to-flux conversion factors for both Pickles stars and GRB models, but in the UV bands their validity range is limited to $b-v=0.1$ and $b-v=0.03$, respectively, 
while BL Lacertae has $b-v \sim 0.8$. 

In order to attenuate possible calibration problems, we thus calculated both the effective wavelengths $\lambda_{\rm eff}$ and count-rate-to-flux conversion factors $\rm CF$ for the UVOT filters by folding the BL Lacertae spectrum with their effective areas \citep[see][]{poo08}. 
We first built a composite observed spectrum of BL Lacertae by combining a mean OM spectrum (obtained from the three XMM-Newton pointings of 2007--2008, \citealt{rai09}) with an average UVOT spectrum (resulting from 16 UVOT observing epochs analysed in this paper). To compensate for the different brightness state, we increased the OM flux densities by $\sim 6$\% so that the two spectra match in the $V$ band. The composite spectrum is shown in Fig.\ \ref{calib} (top panel), together with its log-parabolic fit that we used in the folding procedure.
The resulting effective wavelengths (see Eq.\ 8 in \citealt{poo08}) are:
5439, 4381, 3500, 2776, 2295, and 2225 \AA\ for the $v$, $b$, $u$, $uvw1$, $uvm2$, and $uvw2$ filters, respectively, showing a clear shift towards longer wavelengths in the ultraviolet.
As for the count-rate-to-flux conversion factors, we obtained 2.60, 1.47, 1.65, 4.31, 8.54, and $6.72 \times 10^{-16} \rm \, erg \, cm^{-2} \, s^{-1} \, \AA^{-1}$ from the $v$ to the $uvw2$ band, respectively. These new $\rm CF$ differ from those given by \citet{poo08} for the GRB models by 
$\la 1$\%, with the only exceptions of ${\rm CF}(uvw1)$ and ${\rm CF}(uvw2)$, which are now 8\% larger\footnote{Notice that our ${\rm CF}(uvw1)$ corresponds to that given by \citet{poo08} for the Pickles stars.}.

The new $\lambda_{\rm eff}$ would produce a decrease of extinction in the $uvw1$ and $uvm2$ bands with respect to those given by \citet{poo08}, and an increase in the $uvw2$ band.
Indeed, the Galactic mean extinction curve shows a dramatic bump peaking at $\lambda \sim 2175 \, \AA$ owing to absorption by graphite dust.
Actually, an accurate evaluation of extinction in this critical frequency range requires that the \citet{car89} law is folded with the filter's effective area and BL Lacertae spectrum, similarly to what was done above for the $\lambda_{\rm eff}$ and $\rm CF$:
\begin{equation}
A_{\Lambda}=2.5 \log {{\int {\rm d}\lambda \, E_\Lambda(\lambda) \, F_{\lambda}(\lambda) \, 10^{A(\lambda)/2.5}} \over {\int {\rm d}\lambda \, E_\Lambda(\lambda) \, F_{\lambda}(\lambda)}} \, ,
\end{equation}
where $A_{\Lambda}$ is the extinction in the $\Lambda$ band, $E_\Lambda$ is the effective area of that band, and $F_{\lambda}$ is the source flux density. The result is a Galactic extinction of 1.10, 1.44, 1.74, 2.40, 3.04, and 2.92 mag from the $v$ to the $uvw2$ band, respectively.

We verified the stability of our results by iterating the procedure with the recalibrated UVOT flux densities and $\lambda_{\rm eff}$.
The bottom panel of Fig.\ \ref{calib} shows the mean SED obtained after recalibration of the UVOT data according to our procedure. For comparison, we also show
the OM and UVOT SEDs derived from the average spectra shown in the top panel, for which the amount of Galactic extinction was calculated from the \citet{car89} law at the standard $\lambda_{\rm eff}$. Notice that the recalibration process has shifted $\lambda_{\rm eff}(uvw2)$ redward so much that it overlaps with the standard $\lambda_{\rm eff}(uvm2)$. This is because of the noticeable red tail of the $uvw2$ filter \citep{poo08} as well as to the red BL Lacertae spectrum.
Recalibration has solved the $uvw1$-dip problem, which is a common feature of UVOT SEDs for a number of blazars at different redshifts \citep[see e.g.][]{vil08,rai08b,dam09}. Moreover, it seems to confirm the UV excess claimed by \citet{rai09} that was ascribed to thermal emission from the accretion disc, even if this excess may be less pronounced than indicated by the OM data.
Our analysis highlights the importance of calculating the amount of extinction in the critical UV bands, close to the 2175 \AA\ bump, by folding the Galactic mean extinction law through the effective area curves and source spectrum.
In any case, as pointed out by \citet{fit07}, one has to keep in mind that the use of an average dereddening curve implies a significant error owing to the scatter of Galactic extinction curves.

\subsection{XRT data}
\label{sec_xrt}
The X-ray Telescope (XRT; \citealt{bur05}) data were processed with version 0.12.3 of the {\tt xrtpipeline} task contained in the FTOOLS package, applying standard screening criteria.
Inspection of the light curves revealed that the count rate was low, from 0.15 to 0.24 counts $\rm s^{-1}$, so that observations were performed in photon counting mode, and no correction for pile-up was necessary.
Source and background spectra were extracted with {\tt xselect} from a circular region of 20 pixel (47 arcsec) radius centred on the source and from a surrounding annulus of 30 and 50 pixel radii, respectively.
We used version 011 of the response matrix available in the HEASARC calibration database (CALDB), and calculated the ancillary response file with {\tt xrtmkarf}, using the exposure map created by {\tt xrtpipeline}.
The source spectra were binned with {\tt grppha} to have a minimum of 20 counts in each bin, and they were finally analysed with version 12.5.1 of the {\tt Xspec} task, using the energy channels greater than 0.3 keV.

   \begin{figure}
   \resizebox{\hsize}{!}{\includegraphics[angle=-90]{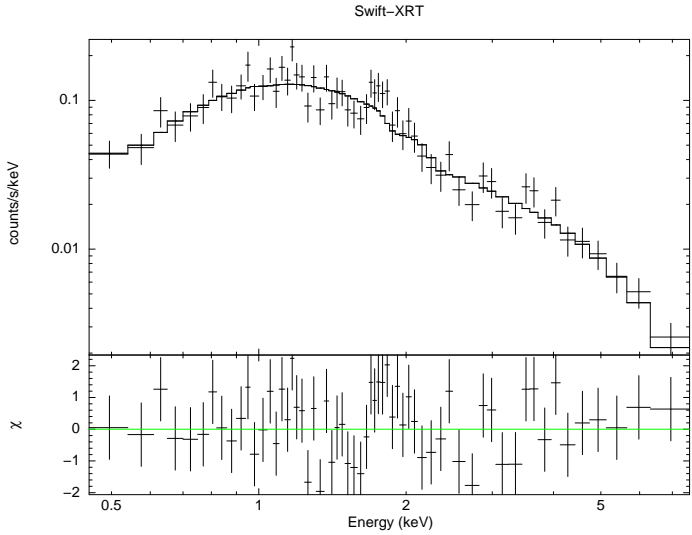}}
      \caption{Swift-XRT spectrum of BL Lacertae on 2008 August 29.
The bottom panel shows the deviations of the observed data from the folded model (a single power law with fixed absorption) in units of standard deviations.} 
         \label{xrt}
   \end{figure}

Spectral analysis was performed for each observation following \citet{rai09}: we first fitted a single power law with free absorption\footnote{We adopted the Tuebingen-Boulder ISM absorption model \citep{wil00}.}, and then fixed the Galactic absorption to $N_{\rm H} = 3.4 \times 10^{21} \, \rm cm^{-2}$, which takes into account both atomic and molecular column density. Statistics is not good enough to evaluate if a double power law model can improve the fit.
The results of spectral fitting on XRT data are reported in Table \ref{xrt_pow} for all observations with an exposure longer than 3 ksec; 
Col.\ 1 gives the date and start time of the observation;
Col.\ 2 its duration;
Col.\ 3 the hydrogen column;
Col.\ 4 the power law photon index;
Col.\ 5 the 1 keV flux density;
Col.\ 6 the $\chi^2/\nu$ (and degrees of freedom).
One spectrum (August 29) is shown in Fig.\ \ref{xrt}. 

Fits with free absorption resulted in a very variable $N_{\rm H}$, which is unlikely to correspond to a real change of absorption. The average and median values are
3.46 and $3.44 \times 10^{21} \, \rm cm^{-2}$, respectively, confirming that the value assumed for the Galactic absorption is quite reasonable. 
Hence, we favoured the second model, whose $\chi^2/\nu$ is usually smaller than in the $N_{\rm H}$-free case, and that produces  results with smaller errors (because of one degree of freedom more). In only two cases (August 23 and September 2) a double power law model with absorption fixed to the Galactic value clearly improved the fit.

The photon index $\Gamma$ ranges from 1.92 to 2.25, indicating a spectrum that oscillates from moderately hard to moderately soft. The average value is 2.07, with standard deviation of 0.08. To understand whether these spectral changes correspond to real variations or are owing to noise, we recall the definition of the mean fractional variation $F_{\rm var}=\sqrt{\sigma^2-\delta^2}/{<f>}$ \citep{pet01}, which is commonly used to characterise variability. Here $<f>$ is the mean value of the variable we are analysing, $\sigma^2$ its variance, and $\delta^2$ the mean square uncertainty. In our case, $\sigma^2=0.006$ is smaller than $\delta^2=0.011$, so that the result is imaginary; thus we conclude that the variations are consistent with noise rather than source variability.

The 1 keV flux density varies between 1.34 and $2.24 \, \mu \rm Jy$, with a mean value of 1.75 and standard deviation of 0.24. In this case $F_{\rm var}=0.11$, and the variations can be considered reliable.

\begin{table*}
\caption{Spectral fitting to the XRT data from Swift observations in 2008. Only exposures longer than 3 ksec are considered.
For each epoch the first line reports the result of the single power law model with free absorption, while the second line shows that obtained when fixing the Galactic absorption to $N_{\rm H} = 3.4 \times 10^{21} \, \rm cm^{-2}$.}             
\label{xrt_pow}      
\centering  
\begin{tabular}{c c c c c c}  
\hline\hline            
Start             & Exp  &  $N_{\rm H}$              & $\Gamma$ & $F_{\rm 1 \, keV}$ & $\chi^2/\nu$ ($\nu$)\\
                  & [s]  & [$10^{21}\, \rm cm^{-2}$] &          & [$\mu$Jy]          &                     \\
\hline
                  &      &                 &                  &                    &           \\
2008-08-20 @ 15:19:01 & 5072 & $3.47^{+0.12}_{-0.11}$ & $2.09^{+0.22}_{-0.20}$ & $1.37^{+0.35}_{-0.27}$ & 0.87 (29) \\
                  &      & 3.4             & $2.07 \pm 0.12$  & $1.34 \pm 0.12$    & 0.84 (30) \\
                  &      &                 &                  &                    &           \\
2008-08-21 @ 11:53:00 & 5200 & $3.12^{+0.09}_{-0.08}$ & $1.98^{+0.18}_{-0.16}$ & $1.59^{+0.30}_{-0.25}$ & 1.03 (39) \\
                  &      & 3.4             & $2.03 \pm 0.10$  & $1.66 \pm 0.13$    & 1.01 (40) \\
                  &      &                 &                  &                    &           \\
2008-08-22 @ 00:45:00 & 5374 & $3.92^{+0.10}_{-0.09}$ & $2.15^{+0.18}_{-0.17}$ & $1.87^{+0.37}_{-0.31}$ & 0.88 (36) \\
                  &      & 3.4             & $2.06 \pm 0.10$  & $1.70 \pm 0.14$    & 0.88 (37) \\
                  &      &                 &                  &                    &           \\
2008-08-23 @ 00:51:01 & 4859 & $1.90^{+0.12}_{-0.11}$ & $1.83^{+0.22}_{-0.21}$ & $1.15^{+0.29}_{-0.23}$ & 1.25 (33) \\
                  &      & 3.4             & $2.06 \pm 0.13$  & $1.50 \pm 0.14$    & 1.34 (34) \\
                  &      &                 &                  &                    &           \\
2008-08-24 @ 13:50:00 & 5118 & $3.30^{+0.10}_{-0.09}$ & $1.95^{+0.17}_{-0.16}$ & $1.76^{+0.34}_{-0.28}$ & 0.96 (41) \\
                  &      & 3.4             & $1.97 \pm 0.10$  & $1.78 \pm 0.14$    & 0.93 (42) \\
                  &      &                 &                  &                    &           \\
2008-08-25 @ 09:07:01 & 5013 & $3.85^{+0.09}_{-0.08}$ & $1.99^{+0.16}_{-0.15}$ & $1.75^{+0.33}_{-0.27}$ & 0.98 (45) \\
                  &      & 3.4             & $1.92 \pm 0.09$  & $1.62 \pm 0.12$    & 0.97 (46) \\
                  &      &                 &                  &                    &           \\
2008-08-26 @ 09:16:00 & 4756 & $3.29^{+0.08}_{-0.07}$ & $1.99^{+0.16}_{-0.15}$ & $1.85^{+0.31}_{-0.27}$ & 0.65 (45) \\
                  &      & 3.4             & $2.01 \pm 0.09$  & $1.89 \pm 0.14$    & 0.64 (46) \\
                  &      &                 &                  &                    &           \\
2008-08-27 @ 00:07:00 & 4197 & $2.81^{+0.10}_{-0.09}$ & $1.99^{+0.21}_{-0.19}$ & $1.69^{+0.37}_{-0.30}$ & 0.62 (37) \\
                  &      & 3.4             & $2.10 \pm 0.11$  & $1.89 \pm 0.15$    & 0.63 (38) \\
                  &      &                 &                  &                    &           \\
2008-08-27 @ 23:59:00 & 3002 & $3.53^{+0.15}_{-0.13}$ & $2.05^{+0.25}_{-0.23}$ & $1.68^{+0.49}_{-0.37}$ & 0.71 (20) \\
                  &      & 3.4             & $2.03 \pm 0.14$  & $1.63 \pm 0.18$    & 0.68 (21) \\
                  &      &                 &                  &                    &           \\
2008-08-29 @ 12:48:00 & 5828 & $3.16^{+0.07}_{-0.06}$ & $2.01^{+0.14}_{-0.13}$ & $1.89^{+0.28}_{-0.24}$ & 1.03 (58) \\
                  &      & 3.4             & $2.05 \pm 0.08$  & $1.97 \pm 0.13$    & 1.02 (59) \\
                  &      &                 &                  &                    &           \\
2008-08-30 @ 08:05:00 & 5044 & $4.69^{+0.09}_{-0.08}$ & $2.38^{+0.18}_{-0.17}$ & $2.86^{+0.53}_{-0.44}$ & 1.01 (50) \\
                  &      & 3.4             & $2.15 \pm 0.08$  & $2.24 \pm 0.14$    & 1.13 (51) \\
                  &      &                 &                  &                    &           \\
2008-08-31 @ 08:11:01 & 5100 & $3.22^{+0.08}_{-0.07}$ & $2.12^{+0.17}_{-0.16}$ & $2.02^{+0.36}_{-0.30}$ & 0.95 (51) \\
                  &      & 3.4             & $2.15 \pm 0.09$  & $2.09 \pm 0.14$    & 0.93 (52) \\
                  &      &                 &                  &                    &           \\
2008-09-01 @ 08:17:01 & 5491 & $3.44^{+0.08}_{-0.07}$ & $1.99 \pm 0.14$ & $1.85^{+0.30}_{0.25}$ & 1.01 (54) \\
                  &      & 3.4             & $1.99 \pm 0.08$  & $1.83 \pm 0.12$    & 0.99 (55) \\
                  &      &                 &                  &                    &           \\
2008-09-02 @ 05:11:01 & 4077 & $3.31^{+0.11}_{-0.10}$ & $2.04^{+0.22}_{-0.20}$ & $1.65^{+0.39}_{-0.31}$ & 1.01 (34) \\
                  &      & 3.4             & $2.06 \pm 0.11$  & $1.67 \pm 0.14$    & 0.98 (35) \\
                  &      &                 &                  &                    &           \\
2008-09-03 @ 14:55:00 & 4791 & $3.63^{+0.08}_{-0.07}$ & $2.11^{+0.18}_{-0.17}$ & $1.92^{+0.36}_{-0.27}$ & 1.01 (44) \\
                  &      & 3.4             & $2.06 \pm 0.10$  & $1.84 \pm 0.13$    & 1.00 (45) \\
                  &      &                 &                  &                    &           \\
2008-09-04 @ 07:06:00 & 5032 & $4.35^{+0.11}_{-0.09}$ & $2.23^{+0.22}_{-0.20}$ & $2.22^{+0.50}_{-0.40}$ & 0.96 (37) \\
                  &      & 3.4             & $2.06 \pm 0.10$  & $1.85 \pm 0.14$    & 1.01 (38) \\
                  &      &                 &                  &                    &           \\
2008-09-05 @ 09:02:00 & 4491 & $4.11^{+0.10}_{-0.09}$ & $2.28^{+0.19}_{-0.18}$ & $2.30^{+0.47}_{-0.38}$ & 1.11 (39) \\
                  &      & 3.4             & $2.16 \pm 0.10$  & $2.02 \pm 0.15$    & 1.12 (40) \\
                  &      &                 &                  &                    &           \\
2008-09-06 @ 05:58:31 & 4676 & $3.21^{+0.11}_{-0.10}$ & $2.22^{+0.22}_{-0.20}$ & $1.58^{+0.38}_{-0.30}$ & 1.16 (32) \\
                  &      & 3.4             & $2.25^{+0.12}_{-0.11}$ & $1.64 \pm 0.14$    & 1.13 (33) \\
                  &      &                 &                  &                    &           \\
2008-09-08 @ 13:46:01 & 4860 & $2.84^{+0.12}_{-0.11}$ & $2.08^{+0.23}_{-0.21}$ & $1.29^{+0.32}_{-0.25}$ & 0.97 (29) \\
                  &      & 3.4             & $2.17 \pm 0.13$  & $1.42 \pm 0.13$    & 0.96 (30) \\
                  &      &                 &                  &                    &           \\
2008-09-09 @ 02:37:00 & 4501 & $4.08^{+0.13}_{-0.12}$ & $2.15^{+0.25}_{-0.23}$ & $1.56^{+0.43}_{-0.33}$ & 1.18 (27) \\
                  &      & 3.4             & $2.04^{+0.13}_{-0.12}$ & $1.38 \pm 0.13$    & 1.17 (28) \\
\hline
\end{tabular}
\end{table*}

\begin{figure}
   \resizebox{\hsize}{!}{\includegraphics{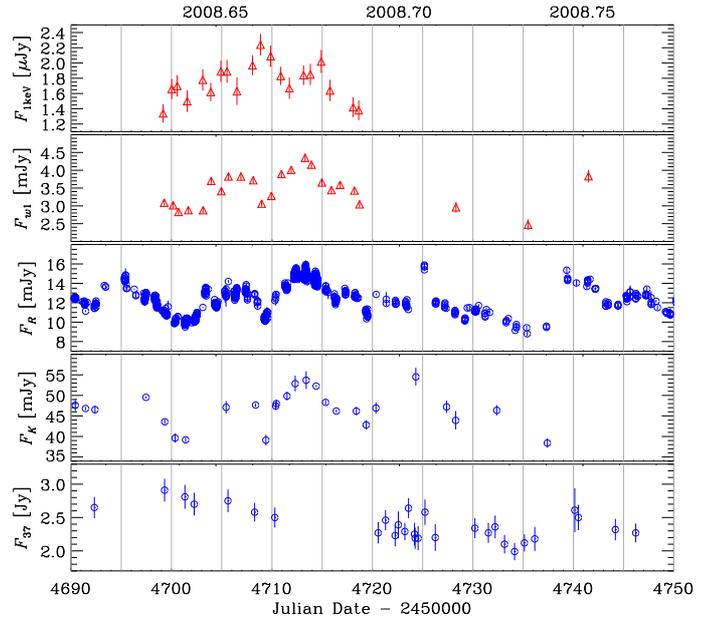}}
      \caption{Details of the multiwavelength behaviour of BL Lacertae in the period around the Swift observations. The X-ray flux density (at 1 keV, top) is compared to that in the UV ($uvw1$ band), optical ($R$ band), near-IR ($K$ band), and radio (37 GHz, bottom) frequency range.} 
         \label{multi}
   \end{figure}

Multiwavelength light curves of BL Lacertae in the period around the Swift observations are shown in Fig.\ \ref{multi}. The source behaviour at 1 keV differs from the common trend characterising the UV, optical, and near-IR bands. In particular, the X-ray flux peaks when the near-IR--UV fluxes reach a minimum. However, there are also similarities, like the flux increase at the beginning of the common observing period, and the final decrease. This may indicate that the 1 keV flux behaviour sometimes is related to the brightness changes that occur at lower wavelengths, while in other cases another variability mechanism prevails.
Indeed, according to \citet{rai09} this frequency domain receives the variable contribution of two different emission components (see also Sect.\ \ref{sec_model}).

\section{XMM-Newton observations}
\label{sec_xmm}

The X-ray Multi-Mirror Mission (XMM) - Newton satellite observed the source during revolution 1545, on 2008 May 16--17, with  a total exposure of $\sim 134 \, \rm ks$. 
Data were processed with the Science Analysis System (SAS) package version 9.0.

\subsection{OM data}
\label{sec_om}
\begin{figure}
   \resizebox{\hsize}{!}{\includegraphics{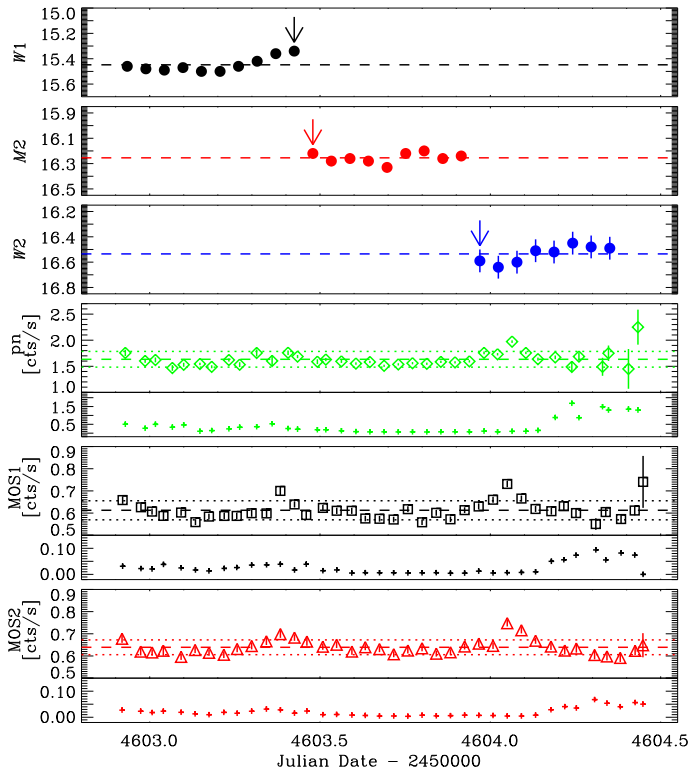}}
      \caption{UV and X-ray light curves obtained by the OM and EPIC instruments onboard XMM-Newton on 2008 May 16--17.
Black, red, and blue filled circles represent UV$W1$, UV$M2$, and UV$W2$ magnitudes, respectively. Arrows mark the data points used for the SED shown in Fig.\ \ref{sed_tot}. 
Green diamonds, black squares, and red triangles refer to pn (0.5--10 keV), MOS1 (0.3--10 keV), and MOS2 (0.3--10 keV) count rates, respectively. Plus signs show the corresponding backgrounds.
In all panels horizontal dashed lines indicate average values and dotted lines represent standard deviations from the mean.} 
         \label{om+epic}
   \end{figure}

The Optical Monitor \citep[OM;][]{mas01} onboard XMM-Newton is a 30-cm telescope carrying six optical/UV filters, and two grisms. BL Lacertae observations in May 2008 consisted of 10 subsequent exposures in UV$W1$, followed by 9 in UV$M2$, and then 8 in UV$W2$. All exposures were $\sim 4000 \, \rm s$ long.
We used the SAS task {\tt omichain} to reduce the data and the tasks {\tt omsource} and {\tt omphotom} to derive the source magnitude. The error on the aperture photometry is 0.03, 0.04, and 0.09 mag for the UV$W1$, UV$M2$, and UV$W2$ filters, respectively.
The resulting light curves are shown in Fig.\ \ref{om+epic}; average magnitudes are UV$W1=15.45$, UV$M2=16.25$, and UV$W2=16.54$.

To obtain flux densities for further analysis, OM magnitudes were corrected for the Galactic extinction calculated according to the \citet[][see Sect.\ \ref{sec_gasp}]{car89} laws at the effective wavelengths of the OM filters (2910, 2310, and 2120 \AA\ for the UV$W1$, UV$M2$, and UV$W2$ filters, respectively). 
Conversion of de-reddened magnitudes into flux densities was done with respect to Vega.

\subsection{EPIC data}

   \begin{figure}
   \resizebox{\hsize}{!}{\includegraphics[angle=-90]{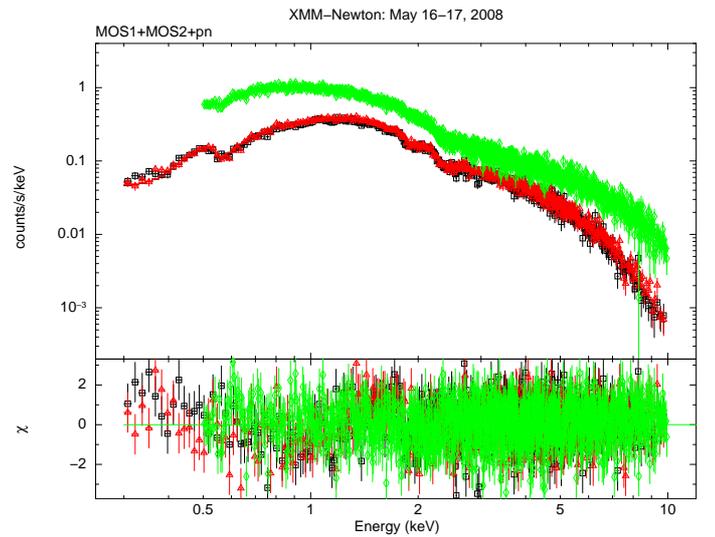}}
      \caption{EPIC spectrum of BL Lacertae on 2008 May 16--17; 
black squares, red triangles, and green diamonds represent MOS1, MOS2, and pn data, respectively.
The bottom panel shows the deviations of the observed data from the folded model (a double power law with fixed absorption) in units of standard deviations.} 
         \label{xmm}
   \end{figure}

The European Photon Imaging Camera (EPIC) onboard XMM-Newton includes three detectors: 
MOS1, MOS2 \citep{tur01}, and pn \citep{str01}.
All instruments were used with a thin filter.
The two MOS cameras observed in small-window imaging mode, while pn was used in timing mode.

We followed the standard prescription to reduce the data, including filtering of high background periods with a threshold of 0.35 counts $\rm s^{-1}$ for MOS, but with a stricter threshold of 0.1 counts $\rm s^{-1}$ for pn.

For both MOS1 and MOS2, we created a filtered sky image, and extracted the source counts from a 50 arcsec radius circular region, while background was evaluated in a circle on an external CCD. 
As for pn, we extracted the source counts from a strip between RAWX=35 and 39, and the background from two strips at columns 24--28 and 48--52.
To get the most reliable and best calibrated events, we used the FLAG==0 selection expression and kept only single and double events (PATTERN$<$=4).
We verified that pile-up effects were not affecting the MOS data with the {\tt epatplot} task.

Through the {\tt grppha} task of the FTOOL package we binned the source spectra to have a minimum of 25 counts in each bin and then analysed them together by means of the {\tt Xspec} task of the XANADU package. Only spectral bins corresponding to energies between 0.3 and 10 keV for MOS1 and MOS2 and in the range 0.5--10 keV for pn were considered, because they have both a better calibration and a higher signal-to-noise ratio. 

The three EPIC spectra were analysed together by first fitting a single power law with free absorption, and then fixing the Galactic absorption to $N_{\rm H} = 3.4 \times 10^{21} \, \rm cm^{-2}$. We also tried a double power law with the same Galactic absorption. The results are shown in Table \ref{fit} (see also Fig.\ \ref{xmm}).
In agreement with \citet{rai09}, the $\chi^2/\nu$ suggests that a single power law with Galactic absorption of $N_{\rm H} = 3.4 \times 10^{21} \, \rm cm^{-2}$ does not represent a good fit to the data. As for the other two fits, the double power law seems to better fit the data, which is also confirmed by a very low F-test probability of $\sim 6.8 \times 10^{-16}$. 
This implies a strong spectral curvature.

To check for possible flux variations, we extracted X-ray light curves from the same source and background regions defined for the spectra, with the same selection expressions. We considered only the events in the time intervals free from high background and belonging to the 0.3--10 keV energy range for MOS1 and MOS2, and 0.5--10 keV for pn.
The source counts were corrected for the background and then binned in one hour intervals. The results are shown in Fig.\ \ref{om+epic}, which also displays the behaviour of the background to check the reliability of the flux variations. 
The background increased significantly only in the last 6 hours. Just before, at $\rm JD \sim 2454604.05$, there is a small flare clearly visible in all three light curves.
Mean source rates for the whole period are 1.63, 0.61, and 0.64 counts $\rm s^{-1}$ for pn, MOS1, and MOS2, respectively, with standard deviations of 0.15, 0.04, and 0.03 counts $\rm s^{-1}$. This means fractional variations $F_{\rm var}$ of 7\%, 6\%, and 4\%, which do not change significantly if we exclude the last 6 hours. 
Hence, we can conclude that the X-ray flux of BL Lacertae is mildly variable on an hour time scale.

\begin{table}
\caption{Spectral fitting to the EPIC (pn+MOS1+MOS2) data from the XMM-Newton observations of 2008 May 16--17.
The first line reports the result of the single power law model with free absorption; the second shows that obtained when fixing the Galactic absorption to $N_{\rm H} = 3.4 \times 10^{21} \, \rm cm^{-2}$, and the third that from a double power law model with the same Galactic absorption.}             
\label{fit}      
\centering  
\begin{tabular}{c c c c}  
\hline\hline            
$N_{\rm H}$               & $\Gamma$ & $F_{\rm 1 \, keV}$ & $\chi^2/\nu$ ($\nu$) \\
$\rm [10^{21} \, cm^{-2}]$  &          &      [$\mu$Jy]     &                    \\
\hline                         
2.60 $\pm$ 0.04        & $1.83 \pm 0.01$ & $1.13 \pm 0.01$   & 1.09 (2030)\\
3.40                   & $1.974 \pm 0.007$ & $1.322 \pm 0.008$ & 1.48 (2031)\\
3.40                   & $2.80^{+0.20}_{-0.17}$, $1.57^{+0.08}_{-0.10}$ &$0.66^{+0.15}_{-0.14}$, $0.67^{+0.13}_{-0.15}$ & 1.05 (2029)\\
\hline
\end{tabular}
\end{table}

   \begin{figure*}
   \sidecaption
   \includegraphics[width=12cm]{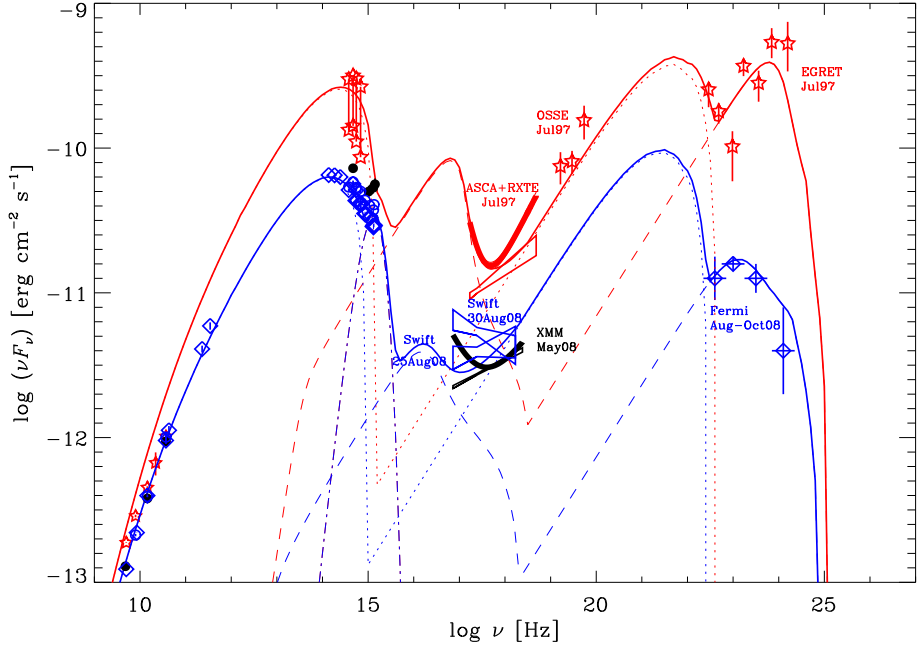}
      \caption{Broad-band SEDs of BL Lacertae in August 2008 (blue) and July 1997 (red). 
The 2008 SED is built with UV and X-ray data from two epochs of Swift observations (Sect.\ \ref{sec_swift}), 2008 August 25 and 30, together with simultaneous optical-to-radio data from the GASP-WEBT collaboration, and contemporaneous $\gamma$-ray data from Fermi \citep[from][]{abd10_sed}. 
The 1997 SED includes EGRET data from \citet{blo97}, OSSE data from the HEASARC archive, ASCA+RXTE spectra from \citet{tan00}, while low-frequency data are from the WEBT archive.
Solid lines represent model fits; we distinguish the low-energy emission component (dotted lines) from the high-energy one (dashed lines); the contribution by an accretion disc of $\sim 16000 \rm \, K$ and $5 \times 10^{44} \rm \, erg \, s^{-1}$ is marked with a dotted-dashed line. We also show in black the SED corresponding to the XMM-Newton observations of May 2008 (Sect.\ \ref{sec_xmm}); both the single power law with free absorption and double power law with Galactic absorption fits to the EPIC spectra are displayed.} 
         \label{sed_tot}
   \end{figure*}

\section{Modelling the SED}
\label{sec_model}

Figure \ref{sed_tot} shows the broad-band SED of BL Lacertae in different brightness states. 
The SED corresponding to 2008 May 16--17 includes the XMM-Newton UV and X-ray data analysed in Sect.\ \ref{sec_xmm}. In order to avoid offsets caused by source variability, the OM spectrum was constructed with the last UV$W1$ data point, the first UV$M2$ (which is close to the last UV$M2$), and the first UV$W2$ datum of the corresponding light curves (see Fig.\ \ref{om+epic}). These data indicate a hard UV spectrum.
Two SEDs in the figure refer to the Swift observations of 2008 August 25 and 30, which were chosen among those analysed in Sect.\ \ref{sec_swift} because of the different spectral slope in the X-ray band. 
The low-frequency part of these three SEDs is built with GASP optical and radio data (see Sect.\ \ref{sec_gasp}); for August 30 near-IR data were also available. 
The strong source variability in the optical band requires that the optical (and near-IR) data are simultaneous with the satellite observations. 
In the radio bands flux variations are slower, so that we used data taken within 2--3 days from the satellite ones, when simultaneous data were not available.
Near-IR, optical, and UV data were corrected for the Galactic extinction; the near-IR and optical data were also corrected for the contribution of the host galaxy (see Sect.\ \ref{sec_gasp}).
In August 2008 the Fermi $\gamma$-ray satellite detected BL Lacertae; the Fermi data we plotted in Fig.\ \ref{sed_tot} were derived from \citet{abd10_sed}.

The August 2008 SEDs indicate a faint, synchrotron-dominated state of the source that we fitted with the rotating helical jet model by \citet[][see also \citealt{rai99}, \citealt{rai03}, \citealt{ost04}]{vil99}.
This model has been used by \citet{rai09} to fit the broad-band SED of BL Lacertae in December 2007 -- January 2008.
Their main finding was that the BL Lacertae broad-band SED cannot be explained by a single synchrotron component plus its self inverse-Compton emission. Indeed, the very strong historical X-ray variability requires an additional synchrotron (plus self inverse-Compton) component. Moreover, the UV excess  suggests thermal contribution from the accretion disc (see Sect.\ \ref{sec_uvot}). 
The SED analysed by \citet{rai09} lacked simultaneous $\gamma$-ray data, which made it impossible to constrain the inverse-Compton emission of the high-energy component. The 2008 August SED in Fig.\ \ref{sed_tot} now offers us the possibility to perform a more detailed analysis.

\begin{table}
\caption{Main parameters of the helical jet model for the fit to both the 2008 faint-state SED and the 1997 outburst-state SED. For each epoch, ``low" and ``high" refer to the low- and high-energy synchrotron plus self inverse-Compton components, respectively. }
\label{modelfit}
\centering
\begin{tabular}{l | r r |  r r}
\hline\hline
          & \multicolumn{2}{c |}{SED 2008} & \multicolumn{2}{c}{SED 1997}\\
Parameter & Low & High & Low & High\\
\hline
$\zeta$                    & $2 \degr$   & $8 \degr$     & $2 \degr$ & $8 \degr$\\
$a$                        & $180 \degr$ & $360 \degr$   & $180 \degr$ & $360 \degr$\\
$\psi$                     & $\bf 5 \degr$   & $\bf 3 \degr$     & $\bf 4.5 \degr$ & $\bf 7 \degr$\\
$\phi$                     & $\bf 150 \degr$ & $\bf 135 \degr$    & $\bf 30 \degr$ & $\bf 70 \degr$\\
$\log \nu'_{\rm s}(0)$     & 14 & 17.8                   & 14 & 17.8\\
$c_{\rm min,max}$          & 2  & 2.5                    & 2 & 2.5\\
$\log l_{\rm min}$         & $-3.2$ & $-3.2$             & $-3.2$ & $-3.2$\\
$\log l_{\rm max}$         & $-1.6$ & $-1.6$             & $-1.6$ & $-1.6$\\
$\log \gamma_{\rm max}(0)$ & 3.7 & 3.8                   & 3.7 & 3.8\\
$c_\gamma$                 & 1 & 1                       & 1 & 1\\
$\log l_\gamma$            &$-0.5$&$-0.5$                & $-0.5$ & $-0.5$\\
$\alpha_0$                 & 0.5  & 0.5                  & 0.5    & 0.5\\
$\Gamma$                   & 7 & 13                      & 7 & 13\\
$c_{\rm s,c}$              & 3 & 1                       & 3 & 1\\
$\log l_{\rm s,c}$         & $-1$ & $-1$                 & $-1$ & $-1$\\
\hline
\end{tabular}
\end{table}

In addition, we also display in Fig.\ \ref{sed_tot} the broad-band SED corresponding to the big outburst of July 1997, which showed a considerable inverse-Compton dominance. 
The X-ray spectra plotted in the figure are the result of the combined analysis of the ASCA and RXTE data by \citet{tan00}. Because of the very strong variability of the source in that period, the authors distinguished between a low state, which was well fitted by a single power law model, and a flare state, for which the best fit was obtained with a double power law model. This last fit resulted in a very strong spectral curvature\footnote{We notice that \citet{tan00} adopted a Galactic total absorption of $N_{\rm H} = 4.6 \times 10^{21} \, \rm cm^{-2}$.}. In July 1997 observations in the $\gamma$-ray band were performed by CGRO. The data from the EGRET instrument onboard CGRO in Fig.\ \ref{sed_tot} were taken from \citet{blo97}, while those from the OSSE detector were derived from the High Energy Astrophysics Science Archive Research Center\footnote{\tt  http://heasarc.gsfc.nasa.gov/} (HEASARC). The low-frequency information is from the WEBT archive; the range of optical flux variation in the period is indicated.
This outburst state of the source was fitted with the same rotating helical model that we used to fit the faint state of 2008. 

In performing the model fits our aim was to see whether it was possible to reproduce the high and low states by changing the geometrical configuration only. Moreover, we took into account the results by \citet{lar10}, who explained the long-term BL Lacertae optical and near-IR variability in terms of variations of the Doppler boosting factor due to changes of the viewing angle of the emitting region.

The resulting model parameters are reported in Table \ref{modelfit}, while the corresponding fits are shown in Fig.\ \ref{sed_tot}. 
We also included blackbody radiation from an accretion disc with a luminosity of $5 \times 10^{44} \rm \, erg \, s^{-1}$ and a temperature of $\sim 16000 \, \rm K$.
The low-energy (radio-to-optical and related inverse-Compton) emission component comes from a helix portion with a pitch angle $\zeta=2\degr$, covering an angle $a=180 \degr$. The maximum Lorentz factor of the relativistic electrons is $\log \gamma_{\rm max}(0)=3.7$, while the bulk Lorentz factor of the plasma in the jet is $\Gamma=7$. 
The high-energy (UV--X-ray and related inverse-Compton) emission component comes from a helix portion with a pitch angle $\zeta=8\degr$, covering an angle $a=360 \degr$. The maximum Lorentz factor of the relativistic electrons is $\log \gamma_{\rm max}(0)=3.8$, while the bulk Lorentz factor is $\Gamma=13$. 
These parameters, as well as those defining the power laws according to which the maximum and minimum emitted frequencies and the flux densities decrease with distance from the jet apex, are maintained fixed.
The difference between the fits to the 1997 and 2008 SEDs is only due to a variation of the orientation of the jet emitting regions, through only two geometric parameters: the angle between the helix axis and the line of sight $\psi$ and the rotation angle $\phi$. The outburst state requires a better alignment of the emitting regions with the line of sight, which implies $\psi$ approaching the helix pitch angle, and smaller rotation angles $\phi$.

\section{Discussion}

The satisfactory fits that we obtained in the previous section for both the outburst and faint states of BL Lacertae give strength to our interpretation of the source SED in terms of two synchrotron plus self inverse-Compton emission components.
In the SED, the synchrotron peak of the low-energy component falls in the near-IR band, and its inverse-Compton reaches a maximum in the 1--50 MeV energy range. The synchrotron and inverse-Compton peaks of the high-energy component occur in the far-UV--soft-X-ray band and in the energy range 0.2--5 GeV, respectively. 
Whether these two components come from two distinct helices or from different regions inside the same helical jet is not clear. We consider it more likely that there is a unique jet, where the high-energy component comes from a region closer to the emitting jet apex than the low-energy one. 
We notice that a double synchrotron component is not an unusual interpretation for the blazar emission properties, as it has been proposed also for Mkn 421 \citep{don09a} and 3C 454.3 \citep{ogl10}.

   \begin{figure}
   \resizebox{\hsize}{!}{\includegraphics{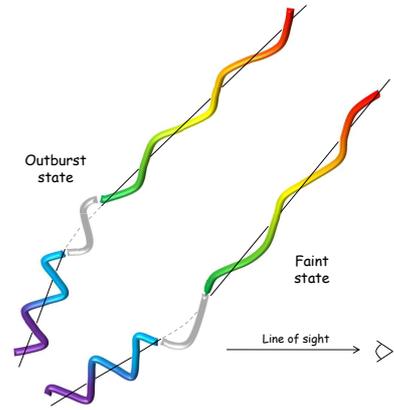}}
      \caption{Sketch of our helical jet model during both the 2008 faint state and the 1997 outburst state.
The angle between the jet axis and the line of sight, $\psi$, has been multiplied by a factor 10 with respect to the values given in Table \ref{modelfit} for clarity.
We distinguish the inner region, emitting the high-energy synchrotron plus self inverse-Compton component (purple-blue) from the outer zone, where the low-energy emission component is produced (green-yellow-red).} 
         \label{sketch}
   \end{figure}

Moreover, the fits show that the whole range of BL Lacertae multiwavelength variability can be interpreted in terms of orientation effects. 
Although the rotating helical jet model we have adopted in the previous section is not a physically complete model, but more a phenomenological approach, it has the advantage of taking into account variations of the orientation of the emitting regions with respect to the line of sight, with consequent changes of the Doppler beaming factor. This is an aspect that is usually neglected by theoretical models of blazar emission, which explain flux and spectral changes uniquely in terms of energetic processes inside the jet.

Our interpretation is in line with previous results. 
\citet{mar08} analysed the evolution of the BL Lacertae optical polarization during 2005, and suggested that the plasma flows along helical streamlines.
According to \citet{vil09a}, the long-term optical and radio behaviour of BL Lacertae suggests a scenario where the emitting plasma flows along a rotating helical path in a curved jet.
This rotating helical structure could be caused by orbital motion in a binary black hole system, coupled with the interaction of the plasma jet with the surrounding medium. Indeed, the binary black hole scenario could explain the periodicity observed in the radio light curves of BL Lacertae \citep{vil04b,vil09a}, the discovery of a precessing jet nozzle with the VLBA \citep{sti03}, and possibly the parsec-to-kiloparsec jet misalignment \citep[see e.g.][]{kha10}.
Moreover, the analysis of the BL Lacertae spectral evolution in 2000--2008 by \citet{lar10} favoured a picture where the optical and near-IR flux and colour variability can be explained by a variable viewing angle of the emitting region. These authors also suggested that a fractal helical structure may be at the origin of the different time scales of variability.

The values of the angles $\zeta=2\degr$ and $\psi=4.5$--$5\degr$, as well as the Lorentz factor $\Gamma=7$ adopted in the model fits for the low-energy component, agree very well with the corresponding values of \citet{lar10}, thus supporting the common interpretation.
The helix pitch angle $\zeta=8\degr$ found for the high-energy component indicates that this inner jet helical region
would be more twisted than the outer, lower-energy one.
In practice, our model results indicate a helical jet whose axis is bent between the X-ray and optical regions by about $2\degr$ (see the values of $\psi$ in Table \ref{modelfit}) and that is more wrapped near the apex and then tends to relax with a decreasing pitch angle. The different orientations assumed by such a jet in 2008 and 1997 are sketched in Fig.\ \ref{sketch}, where all the angles are strongly increased for clarity.

The thermal emission component that we added to the helical jet model is justified by the UV excess found in the OM data from XMM-Newton, and (though with less evidence) in the UVOT data from Swift. As discussed in the present paper and in \citet{rai09}, the amount of this excess strongly depends on both the Galactic extinction and instrument calibration, but it is not easy to cancel it out completely.
In any case, \citet{cap10} showed that after twelve years from the first detection of the H$\alpha$ broad emission line by \citet{ver95} and \citet[][see also \citealt{cor00}]{cor96}, the H$\alpha$ line is still there, even more luminous than before. This suggests that a disc is also there to photoionise the broad line region.

Photons coming from the disc or broad line region could then enter the jet, and be inverse-Compton scattered, giving rise to other high-energy emission components that are sometimes invoked to account for the SED properties of blazars.
In particular, the 1997 outburst state has previously been interpreted by \citet{mad99} in terms of three emission components: synchrotron, synchrotron self-Compton, and Comptonisation of the broad emission line flux.
Similar results were obtained by \citet{boe00} and by \citet{rav02}.
Our ``geometrical" interpretation does not require these external-Compton emission components, which are not expected to contribute if the jet emission regions are parsecs away from the central black hole \citep[see e.g.][]{sik08,mar10,abd10_nature}.

\begin{acknowledgements}
We acknowledge Ann E. Wehrle for useful comments.
This research has made use of NASA's Astrophysics Data System.
The Torino and Palermo teams acknowledge financial support by the Italian Space Agency through contract ASI-INAF I/088/06/0 for the Study of High-Energy Astrophysics.
The Abastumani Observatory team acknowledges financial support by the
Georgian National Science Foundation through grant GNSF/ST08/4-404.
St.Petersburg University team acknowledges support from Russian RFBR foundation via grant 09-02-00092.
AZT-24 observations are made within an agreement between  Pulkovo, Rome and Teramo observatories.
This work was partly supported by the Taiwan National Science Council grant No. 96-2811-M-008-033.
This paper is partly based on observations carried out at the German-Spanish Calar Alto Observatory, which is jointly operated by the MPIA and the IAA-CSIC. Acquisition of the MAPCAT data is supported in part by the Spanish Ministry of Science and Innovation and the Regional Government of Andaluc\'{\i}a through grants AYA2007-67626-C03-03 and P09-FQM-4784, respectively.
The Submillimeter Array is a joint project between the Smithsonian Astrophysical Observatory and the Academia Sinica Institute of Astronomy and Astrophysics and is funded by the Smithsonian Institution and the Academia Sinica.
The BU team  acknowledges  financial support  by NASA through Fermi Guest Investigator grants NNX08AV65G and NNX08AV61G and by the NSF through grant AST-0907893.
The Mets\"ahovi team acknowledges the support from the Academy of Finland
to our observing projects (numbers 212656, 210338, and others).
\end{acknowledgements}

\end{document}